\documentclass[sigconf, nonacm]{acmart}

\usepackage{amsmath}
\usepackage{mathtools}
\usepackage{enumitem}

\AtBeginDocument{%
  }

\setcopyright{none} 
\begin{document}


\title{\textsf{WaveTune}: Wave-aware Bilinear Modeling for Efficient GPU Kernel Auto-tuning}

%%%%%% -- PAPER CONTENT STARTS-- %%%%%%%%
\begin{abstract}
The rapid adoption of Large Language Models (LLMs) has made GPU inference efficiency an increasingly critical system concern. 
The runtime of LLM workloads is largely dominated by tile-based kernels, particularly General Matrix Multiplications (GEMMs). 
Although these kernels are highly optimized, their performance remains sensitive to a large space of runtime parameters, such as tile sizes and pipeline stages. 
The interaction between these parameters and hardware resources leads to a non-convex optimization landscape. 
Existing approaches to parameter configuration—including search-based auto-tuning, heuristic rules, and learned cost models—face a fundamental trade-off between performance optimality and runtime efficiency.

In this paper, we present \textsf{WaveTune}, a wave-aware framework for runtime kernel auto-tuning. First, we introduce a unified mapping method to handle input diversity and decompose the configuration space to manage high dimensionality. 
Second, we develop an analytical wave-aware bilinear model that accurately predicts kernel latency. 
Third, we design a sparse sampling scheme based on wave structures and a lightweight dual-table retrieval mechanism to minimize runtime overhead. As a result, \textsf{WaveTune} enables precise and efficient runtime configuration for GPU kernels.
Across three representative kernels and five GPU architectures, \textsf{WaveTune} consistently achieves near-optimal kernel performance, delivering up to \textbf{1.83$\times$} kernel-level speedup and up to \textbf{1.33$\times$} end-to-end TTFT reduction, while reducing runtime decision overhead by \textbf{five orders of magnitude} compared to exhaustive search.
These results demonstrate that \textsf{WaveTune} effectively eliminates the traditional trade-off between configuration latency and execution optimality, providing a practical and robust solution for high-performance LLM inference. 
%The source code is available\footnote{\url{https://anonymous.4open.science/r/logos-EE78/.}}.

\end{abstract}


\author{Kaixuan Zhang}
\email{zks1anx@sjtu.edu.cn}
\affiliation{%
  \institution{Shanghai Jiao Tong University}
  \city{Shanghai}
  \country{China}
}

\author{Chutong Ding}
\email{dingchutong2004@163.com}
\affiliation{%
  \institution{Shanghai Jiao Tong University}
  \city{Shanghai}
  \country{China}
}

\author{Shiyou Qian}
\authornotemark[1]
\email{qshiyou@sjtu.edu.cn}
\affiliation{%
  \institution{Shanghai Jiao Tong University}
  \city{Shanghai}
  \country{China}
}

\author{Luping Wang}
\authornotemark[1]
\email{chamu.wlp@alibaba-inc.com}
\affiliation{%
  \institution{Alibaba Group}
  \city{Hangzhou}
  \country{China}
}

\author{Jian Cao}
\email{cao-jian@sjtu.edu.cn}
\affiliation{%
  \institution{Shanghai Jiao Tong University}
  \city{Shanghai}
  \country{China}
}

\author{Guangtao Xue}
\email{gt_xue@sjtu.edu.cn}
\affiliation{%
  \institution{Shanghai Jiao Tong University}
  \city{Shanghai}
  \country{China}
}

\author{Cheng Huang}
\email{xiaoluo.hc@alibaba-inc.com}
\affiliation{%
  \institution{Alibaba Group}
  \city{Hangzhou}
  \country{China}
}

\author{Guodong Yang}
\email{luren.ygd@taobao.com}
\affiliation{%
  \institution{Alibaba Group}
  \city{Hangzhou}
  \country{China}
}

\author{Liping Zhang}
\email{liping.z@alibaba-inc.com}
\affiliation{%
  \institution{Alibaba Group}
  \city{Hangzhou}
  \country{China}
}

\renewcommand{\shortauthors}{Zhang et al.}

\maketitle

\renewcommand\thefootnote{\fnsymbol{footnote}}

\footnotetext[1]{Corresponding authors}

\section{Introduction}
\label{sec:intro}

The rapid scaling and deployment of LLMs has made GPU inference efficiency a critical bottleneck. 
In practice, GPU compute cost is dominated by a small set of kernels, notably matrix multiplications and attention~\cite{sarathi,synperf,simai}.
Although the underlying algorithms of these kernels are well established and widely implemented in modern GPU libraries and frameworks~\cite{cublas,deepgemm,fa2}, their performance remains highly sensitive to a large number of runtime configuration parameters, including tile sizes, the number of warps, pipeline stages, and swizzling strategies.

Selecting an optimal config for kernels is fundamentally challenging due to the high-dimensional and hardware-dependent nature of the search space. 
The interaction between configuration parameters and GPU resources (e.g., registers, shared memory, and SM occupancy) leads to complex and non-linear performance behavior. 
As a result, suboptimal configs can significantly degrade performance, especially under dynamic input shapes in real-world LLM serving workloads \cite{triton_attn, gpuportability}.

Existing approaches to kernel configuration have been extensively studied in both industry and academia, which can be broadly categorized into three classes: search-based auto-tuning, rule-based heuristics, and data-driven cost models. Search-based methods explore large configuration spaces to identify near-optimal configurations, but suffer from exponential complexity that prevents comprehensive offline coverage and incurs prohibitive latency when applied online~\cite{gpuportability}. Rule-based heuristics enable efficient runtime decisions with negligible overhead, yet often lack robustness across diverse workloads and hardware architectures, leading to suboptimal performance~\cite{vllm_git,triton_attn,sglang_git}. Data-driven cost models (e.g., gradient-boosted trees~\cite{autotvm,xgboost,ansor}) predict kernel performance to guide search, but are primarily designed for offline tuning and introduce non-negligible configuration overhead, making them ill-suited for latency-sensitive serving scenarios. Overall, these approaches remain constrained by an inherent tension between achieving high performance and maintaining low runtime overhead.

In this work, we argue that GPU kernel performance is not an unstructured black-box function, but instead exhibits a strong physical structure governed by hardware execution dynamics. 
Specifically, we observe that kernel latency follows a \textit{wave-conditioned piecewise bilinear pattern}, where discrete execution waves introduce structured discontinuities, while intra-wave behavior remains approximately linear. 

Based on this insight, we propose \textsf{WaveTune}, a lightweight framework for runtime kernel auto-tuning. 
\textsf{WaveTune} introduces a unified mapping of diverse kernel inputs and a decomposition of the configuration space to manage input heterogeneity and high dimensionality. It constructs an accurate wave-aware piecewise bilinear model for kernel latency prediction, capturing the interplay between workload structure and hardware execution dynamics. 
% It further designs an efficient dual-table-based runtime lookup mechanism that replaces expensive online search with constant-time configuration selection. 
At runtime, the system predicts latency using precomputed model coefficients and selects the optimal config via lightweight table-based retrieval, incurring only microsecond-level overhead.

We comprehensively evaluated \textsf{WaveTune} on three representative kernels under diverse input workloads, spanning five modern GPU architectures from two major vendors.
At the kernel level, it achieves up to \textbf{1.83$\times$} speedup over default heuristics, consistently approaching the performance of exhaustive search. 
When integrated into an LLM serving system, it delivers up to \textbf{1.33$\times$} end-to-end Time-To-First-Token (TTFT) reduction, while maintaining negligible runtime overhead.

In summary, this paper makes the following contributions:
\begin{itemize}
    \item We identify a wave-conditioned piecewise bilinear structure in GPU kernel latency, revealing a strong physical prior that unifies discrete wave effects and continuous intra-wave scaling.

    \item We propose \textsf{WaveTune}, a runtime kernel auto-tuning framework that integrates sparse profiling and wave-aware bilinear modeling, enabling efficient and deterministic selection via lightweight model-based evaluation.

    \item We demonstrate that \textsf{WaveTune} generalizes across diverse kernels, input workloads, and GPU architectures, consistently achieving near-optimal performance with microsecond level overhead and delivering significant end-to-end latency reductions when integrated into LLM inference systems.
\end{itemize}

\section{Background: Tile-based Kernel Execution}
\label{sec:background}

% \subsection{Tile-based Kernel Execution Model in LLMs}
% \label{subsec:tile_execution}

Modern LLM inference workloads are dominated by compute-intensive kernels such as GEMM and FlashAttention \cite{fa}. To achieve high throughput, these kernels adopt \textit{tile-based} execution, decomposing computation into thread-block-level work units to maximize data locality and parallelism. These thread blocks are then scheduled and executed following the GPU execution model.

Modern GPUs execute a kernel by launching a grid of thread blocks, where each block is independently scheduled onto a Streaming Multiprocessor (SM). An SM can host only a limited number of concurrent blocks, strictly bounded by hardware resource limits such as registers and shared memory~\cite{nvidia_cudaguide, cutlass_docs}.
When the total number of thread blocks exceeds the total concurrent capacity of the GPU's, the execution is divided into multiple rounds, called \textit{waves} (Figure~\ref{fig:wave_quantization}). Since new blocks can only be scheduled after resources are released by completed ones, the total execution time becomes sensitive to the number of waves. As the grid size crosses the hardware capacity boundary, additional waves may be required, introducing discontinuities in execution behavior~\cite{nvidia_wave_quantization}.

Optimizing these tile-based kernels exposes a massive, high-dimensional configuration space. Beyond basic tile sizes, performance is governed by tightly coupled execution parameters such as warp-level parallelism, software pipeline stages, and memory access patterns, which jointly determine workload decomposition and hardware resource utilization~\cite{gpuportability,triton_attn}.

Crucially, selecting an optimal kernel config involves a complex physical trade-off. Increasing tile sizes or intra-block parallelism (e.g., number of warps or pipeline stages) may improve computational efficiency, but it simultaneously increases per-block resource consumption (e.g., registers and shared memory), which can reduce the number of concurrently resident blocks per SM due to hardware resource constraints. This interplay induces highly input- and hardware-dependent execution dynamics, making efficient runtime kernel configuration a central challenge in LLM inference~\cite{cutlass_docs,nvidia_cudaguide,tritonblas}.

\section{Key Observations and Motivation}
\label{obs_motivation}

% This requirement for adaptivity is particularly critical in \textit{LLM inference}. 
% Inference workloads exhibit significant variability in per-request computation demand due to differences in both input and generated sequence lengths, which directly affect latency and resource utilization characteristics and are known only at runtime~\cite{vllm,orca,sarathi}.

\subsection{Runtime Kernel Configuration Challenges}
\label{obs:autotune_challenge}

The selection of optimal GPU kernel configs in LLM inference workloads requires navigating a massive combinatorial design space shaped by the interaction of three primary factors:
\textit{input workload characteristics} (e.g., dynamic variations in batch size and sequence length),
\textit{kernel implementation diversity} (ranging from distinct operators to alternative implementation variants),
and the \textit{target hardware architecture} (e.g., NVIDIA B200 or AMD MI355X).

Different configuration strategies impose varying trade-offs among resource occupancy, instruction throughput, and latency hiding capability.
Consequently, configs that perform well under one specific setting often become suboptimal when workloads or hardware change, motivating the need for adaptive runtime selection strategies~\cite{gpuportability,hipautotune,benchmarking_autotuning,triton_attn}.
However, as summarized in Table~\ref{tab:design_space}, conventional configuration methodologies—including offline search~\cite{hipautotune,benchmarking_autotuning,autotvm}, expert heuristics~\cite{deepgemm,sglang_git,vllm_git,flashinfer}, and machine learning (ML) cost models~\cite{autotvm,xgboost,ansor,cutlass_tailor}—struggle to address this dynamism effectively, with each approach exhibiting distinct limitations across accuracy, generalization, and runtime efficiency.

\begin{table}[t]
\vspace{-0.5em}
\centering
\caption{Comparison of kernel configuration strategies.}
\vspace{-3mm}
\label{tab:design_space}
\fontsize{7}{8.4}\selectfont
\begin{tabular}{lccc}
\toprule
\textbf{Method} & \textbf{Accuracy} & \textbf{Generalization} & \textbf{Runtime Efficiency} \\
\midrule
Offline Search & High & Low & Low \\
Expert Heuristics & Low & Low & High \\
ML Cost Model & Medium & Medium & Low \\
\midrule
\textbf{\textsf{WaveTune(ours)}} & \textbf{High} & \textbf{High} & \textbf{High} \\
\bottomrule
\end{tabular}
\vspace{-4mm}
\end{table}

\noindent\textbf{Infeasibility of Offline Search.} 
Pre-computing optimal configs for the entire input space is intractable: while exhaustive configuration search can achieve high accuracy for a fixed input  on a given hardware~\cite{triton_attn,gpuportability}, it suffers from poor generalization beyond pre-tuned settings. Consequently, when confronted with dynamic, unseen shapes at deployment, it either necessitates prohibitive online tuning or falls back to suboptimal defaults, resulting in low runtime efficiency or degraded performance. 
The combinatorial explosion of kernel input dimensions renders exhaustive precomputation impractical, and scaling this approach across diverse kernels and evolving hardware further amplifies its inefficiency.

\noindent\textbf{Rigidity of Expert Heuristics.}
Hand-tuned heuristics used in high-performance libraries offer high runtime efficiency due to their lightweight nature~\cite{deepgemm,flashinfer,fa2}, but suffer from low accuracy and poor generalization. They rely on simplified assumptions and manual calibration that fail to capture complex, non-linear execution dynamics across inputs, kernels, and hardware. 
These rules require significant engineering effort and often produce suboptimal throughput when deployed beyond their intended regimes, reflecting their limited ability to generalize across diverse workloads and hardware.

\noindent\textbf{High Runtime Overhead of ML Cost Models.} 
Data-driven ML cost models can achieve moderate accuracy and some degree of generalization across configs~\cite{cutlass_tailor}, but incur low runtime efficiency when deployed for online selection due to non-negligible inference overhead on the critical path (Section~\ref{sec:e2e_speedup} and~\ref{sec:overhead}). 
Their reliance on online feature extraction and iterative evaluation often leads to sub-millisecond to millisecond latency on the critical path, which can rival or even exceed kernel execution time in LLM serving workloads~\cite{orca,fa,sarathi,vllm}, making them unsuitable for latency-sensitive production systems where configuration decisions must incur near-zero overhead.

\vspace{0.5em}
\noindent\textit{These limitations highlight a fundamental gap in the design space: existing approaches fail to simultaneously achieve high accuracy, strong generalization, and near-zero runtime overhead (Table~\ref{tab:design_space}).
Resolving this tension requires a fundamentally new paradigm---a lightweight analytical model that approaches the accuracy of exhaustive search while offering the microsecond speed of basic heuristics, as realized by \textsf{WaveTune}.}

\subsection{Empirical Analysis of Real-World Wave Dynamics}
\label{obs:wave_dynamics}

To understand the fundamental performance characteristics of modern GPU kernels and expose the limitations of existing performance modeling methodologies, we conducted a large-scale auto-tuning campaign across several representative tile-based kernels, including Dense GEMM, Grouped GEMM (MoE), and FlashAttention~\cite{deepgemm,sglang_git,triton_attn}. Without loss of generality, we focus on the standard BF16 GEMM kernel from the \texttt{DeepGEMM} library~\cite{deepgemm} running on an NVIDIA H100 GPU, with performance data collected using the PyTorch Profiler~\cite{torch_profiler}. We adopt \texttt{DeepGEMM} as our primary testbed because tile-based GEMM constitutes the computational backbone of modern deep learning workloads~\cite{accelerating}, the library achieves state-of-the-art performance competitive with proprietary vendor implementations~\cite{deepgemm}, and its open-source design exposes a rich configuration space of approximately 3{,}000 candidates spanning more than a dozen tunable parameters---such as tile sizes, pipeline stages, and warp configurations---that remain inaccessible in black-box libraries like cuBLAS~\cite{cublas}.

\begin{figure}[t]
    \centering
    \includegraphics[width=0.75\linewidth]{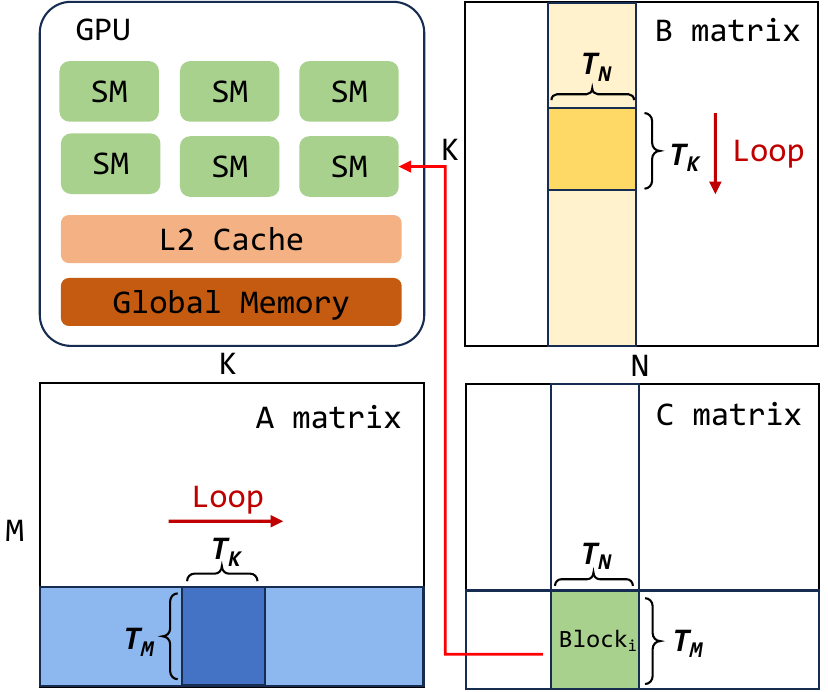} 
    \caption{Illustration of the Tiled GEMM execution model.}
    \label{fig:gemm_execution}
\end{figure}

As illustrated in Figure~\ref{fig:gemm_execution}, a standard GEMM kernel computes a matrix multiplication problem of size $(M, N, K)$, i.e., multiplying $A \in \mathbb{R}^{M \times K}$ and $B \in \mathbb{R}^{K \times N}$, by spatially partitioning the result matrix $C \in \mathbb{R}^{M \times N}$ into a logical grid of thread blocks.
Under a tile config $(T_M, T_N, T_K)$, the total number of launched thread blocks is $G = \lceil M/T_M \rceil \times \lceil N/T_N \rceil$, where each thread block is scheduled onto one of the available SMs and computes a tile of size $T_M \times T_N$ of $C$.
Within each thread block, the kernel iteratively processes the reduction dimension $K$ using a tiled loop.
Specifically, the computation proceeds in $L = \lceil K / T_K \rceil$ iterations, where each iteration loads a pair of tiles of size $T_M \times T_K$ and $T_K \times T_N$ from the input matrices and accumulates their partial products into the corresponding tile of $C$.

% As illustrated in Figure~\ref{fig:gemm_execution}, a standard GEMM kernel computes a matrix multiplication problem of size $(M, N, K)$ by spatially partitioning the output matrix $C \in \mathbb{R}^{M \times N}$ into a logical grid of thread blocks.
% Each thread block is scheduled to execute on one of the available SMs and is responsible for computing a tile of size $T_M \times T_N$ of the output matrix.
% Under a tile configuration $(T_M, T_N, T_K)$, the total number of launched thread blocks is $G = \lceil M/T_M \rceil \times \lceil N/T_N \rceil$.
% Within each thread block, the kernel iteratively processes the reduction dimension $K$ using a tiled loop.
% Specifically, the computation proceeds in $L = \lceil K / T_K \rceil$ iterations, where each iteration loads a pair of tiles of size $T_M \times T_K$ and $T_K \times T_N$ from the input matrices and accumulates their partial products into the output tile.

It is important to note that for \textit{persistent kernels}—a prevalent design pattern on modern architectures like NVIDIA Hopper~\cite{nvidia_hopper_whitepaper}—the physically launched grid size is often clamped to the number of available SMs to maximize residency.
In such designs, $G$ represents the total set of logical tasks organized as a global work queue, from which resident worker blocks dynamically fetch and process tasks.

\begin{figure}[t]
\centering
\includegraphics[width=0.7\linewidth]{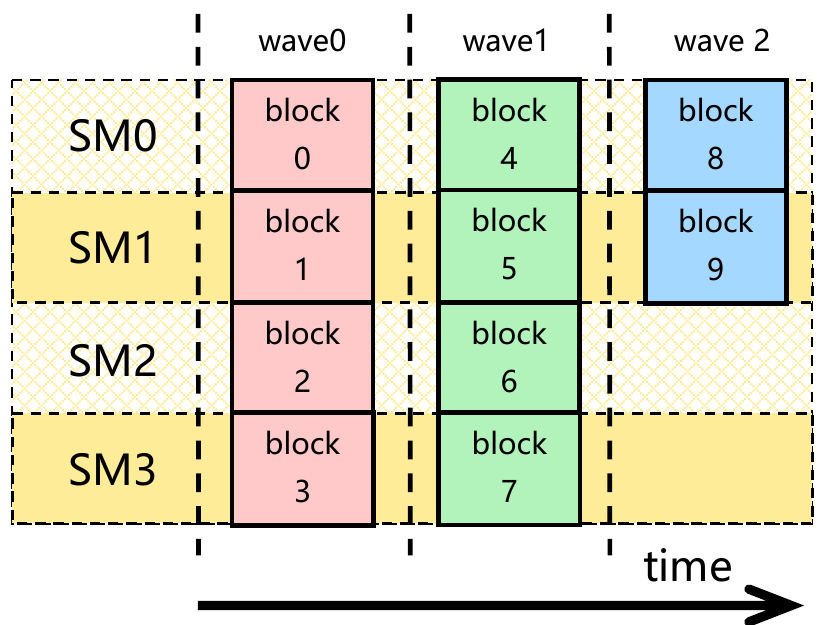}
\caption{Illustration of the wave quantization effect. A total of 10 thread blocks are scheduled across 4 SMs. The execution spans 3 waves: the first 2 waves are fully occupied, while the final quantized wave only partially utilizes the GPU, leaving two SMs idle.}
\label{fig:wave_quantization}
\end{figure}

% Based on these logical definitions, we analyze the complex interplay between software configurations and hardware execution dynamics through systematic profiling. 

Existing performance models can be broadly categorized into two paradigms. The first adopts a \textbf{step} function approximation of latency~\cite{tritonblas,neusight}:
$T_{\text{total}} \approx T_{\text{wave}} \cdot \lceil G / N_{SM} \rceil$.
Here, $G$ denotes the total number of thread blocks in the grid, and
$N_{SM}$ represents the number of SMs available on the GPU.
The term $\lceil G / N_{SM} \rceil$ therefore corresponds to the number of waves required to execute all blocks.
$T_{\text{wave}}$ denotes the latency of one wave, i.e., the
time required for a batch of thread blocks that can simultaneously occupy
all SMs to complete execution.
This formulation is rooted in \textit{wave quantization}~\cite{nvidia_wave_quantization,cutlass_docs}: because the number of SMs is finite, thread blocks are issued in groups whose size is bounded by available resources. When the total number of blocks is not an exact multiple of the maximum number of concurrently resident blocks, the final wave exhibits reduced occupancy, leading to hardware underutilization. As a result, kernel execution is abstracted as a sequence of discrete waves with an approximately constant per-wave cost (Figure~\ref{fig:wave_quantization}).
In contrast, the second paradigm adopts a purely \textbf{linear} perspective~\cite{linear_model,roofline}, modeling latency as a direct function of the aggregate workload volume—measured in total FLOPs or memory bytes, treating the GPU as a continuous throughput device and thus obscuring the discrete impact of wave quantization.

\begin{figure*}[t]
    \centering
    \includegraphics[width=\textwidth]{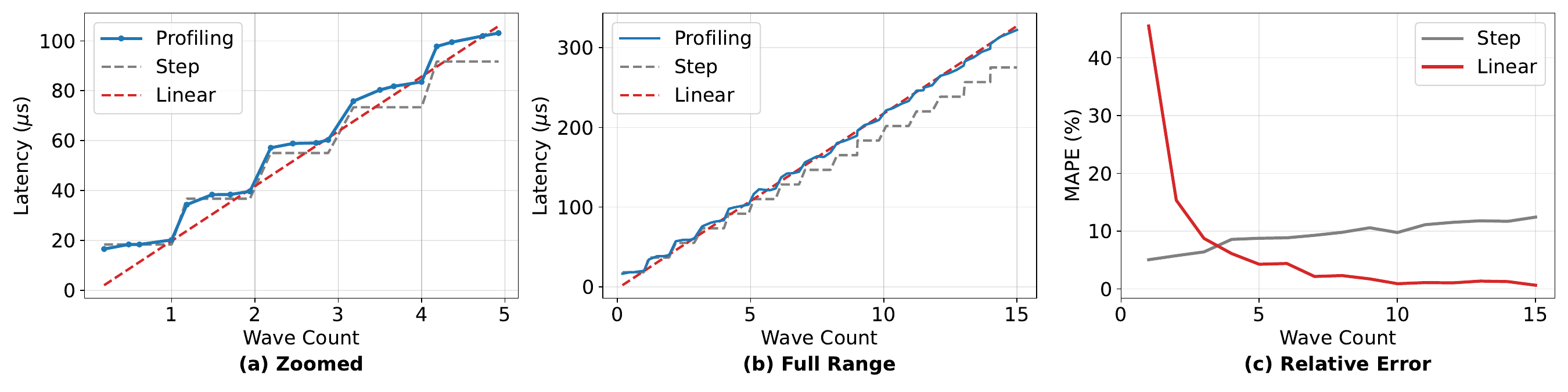}
    \vspace{-8mm}
    % \caption{Profiling results on an H100 (132 SMs) for a GEMM kernel with fixed $K=4096$ and tile configuration $(T_M, T_N, T_K) = (128, 64, 64)$. 
    % The problem size varies in $M$ and $N$, which changes the total number of thread blocks $G$ while keeping the per-block computation constant. 
    % The horizontal axis denotes the wave count $w$. 
    % (a) A zoomed-in view contrasting measured latency (blue) with Step (grey) and Linear (red) models at small grid sizes. 
    % (b) A full-range view showing the evolution toward asymptotic linearity. 
    % (c) Relative error (MAPE) comparison demonstrating the crossover in model fidelity as workload scales.}
    \caption{Profiling results on an H100 (132 SMs) for a GEMM kernel with fixed $K=4096$ and tile configuration $(T_M, T_N, T_K) = (128, 64, 64)$. Varying $M$ and $N$ changes the number of thread blocks $G$ while keeping per-block computation constant; the x-axis denotes the wave count $w$. (a) Zoom-in comparing measured latency (blue) with Step (grey) and Linear (red) models at small grids. (b) Full-range view showing asymptotic linearity. (c) Relative error (MAPE), highlighting the crossover in model fidelity.}

    \label{fig:wave_dynamics}
    \vspace{0mm}
\end{figure*}

However, our systematic profiling shows that neither the rigid step function nor the simplified linear view adequately captures the complex execution dynamics of modern GPU kernels. As shown in Figure~\ref{fig:wave_dynamics}, real-world performance transcends these simplified views, exhibiting complex dynamics that evolve from step-like to linear as the workload scales.
In the initial phase (Waves 1--3), the latency profile aligns relatively well with the rigid step model, exhibiting distinct quantization boundaries. 
However, as the number of waves increases, a linear growth trend emerges within each wave, gradually smoothing the step-like structure until the overall profile converges to a linear trajectory.

This evolution is quantitatively reflected in the model accuracy (Figure~\ref{fig:wave_dynamics}c). 
The linear model, while failing catastrophically in the small-wave regime with errors exceeding 40\%, rapidly improves as the workload scales, eventually achieving high fidelity ($<2\%$ error).
Conversely, the step model exhibits the opposite behavior: it maintains low error at small grid sizes but gradually accumulates deviation as the workload increases.

To understand the root causes of the observed transition from a step-function to a linear regime, we analyze the interplay between workload scale and execution non-determinism. We identify two primary mechanisms that drive this convergence:

\noindent\textbf{Asymptotic Amortization of Quantization Overhead.}
The ``step'' effect in the performance profile arises from the tail wave—the final batch of blocks that may not fully occupy the GPU. 
However, the magnitude of this quantization penalty is bounded by the duration of a single wave ($T_{wave}$). 
As the problem size increases, the total execution time ($T_{total}$) grows linearly. 
Consequently, the relative impact of the fixed quantization penalty ($T_{wave}/T_{total}$) diminishes asymptotically. 
For large-scale workloads, the performance curve naturally aligns with the linear throughput baseline, rendering the discrete steps negligible.

\noindent\textbf{Desynchronization via Execution Variance and Greedy Scheduling.}
The idealized step model premises that all thread blocks within a wave exhibit identical latency, implying that waves complete atomically.
In practice, block execution latencies are stochastic and arise from multiple sources.
These include variability induced by the memory hierarchy—such as cache hit/miss effects, bank conflicts, memory access coalescing inefficiencies, and contention in shared cache and DRAM resources~\cite{linear_model,memeory_dissecting}—as well as instruction- and execution-level effects, including warp scheduling decisions, pipeline stalls due to data dependencies or control divergence, and interference among concurrently resident blocks or kernels competing for shared on-chip and off-chip resources~\cite{warp_scheduling,warp_scheduling_1,gpu_survey}. At the same time, modern dispatch mechanisms—whether the hardware-managed GigaThread Engine \cite{nvidia_cudaguide,li2017cta,song2016ctascheduling,zhang2017tlpcta} or a software-managed persistent work queue
\cite{cutlass_docs,colfax_persistent_kernel}—operate \textit{greedily}: as soon as an SM completes a task, it immediately fetches the next available task, independent of its peers.

\begin{figure*}[t]
    \centering
    \includegraphics[width=\textwidth]{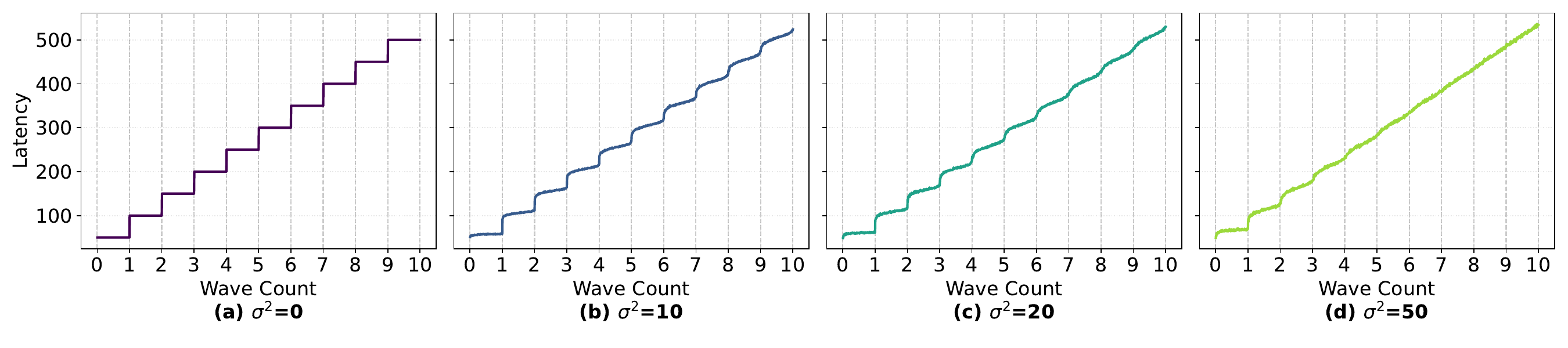}
    \vspace{-8mm}
    \caption{Discrete-event simulations of 132 SMs with a fixed mean block latency $\mu=50$ (arbitrary time units) and varying variance $\sigma^2$.
    As variance increases from 0 to 50, the accumulated desynchronization among SMs progressively erodes the rigid wave boundaries, transforming the step-like profile into a continuous linear trend.}
    \label{fig:wave_simulation}
\end{figure*}

This combination of latency variance and greedy scheduling erodes the rigid execution-time boundaries implied by the wave abstraction.
As a proof-of-concept, we performed a discrete-event simulation of a parallel dispatched workload where task durations follow a Gaussian distribution, systematically sweeping the standard deviation ($\sigma$) around a fixed mean ($\mu$).
As visualized in Figure~\ref{fig:wave_simulation}, the results capture the progressive breakdown of the wave boundaries: while the zero-variance baseline exhibits perfect steps, increasing variance causes the execution timelines of parallel SMs to \textit{desynchronize}. This effect smooths out the discrete steps, reproducing the asymptotic linearity observed in Figure~\ref{fig:wave_dynamics}.

\vspace{0.5em}
\noindent\textit{These observations expose a fundamental limitation of the classical wave abstraction: GPU kernel latency is neither purely discrete nor purely continuous, but a hybrid effect shaped by quantization and variance. This motivates the need for a unified formulation that captures both step-like behavior in the small-wave regime and linear scaling in the large-wave limit.}

\begin{figure}[t]
    \centering
    \includegraphics[width=\linewidth]{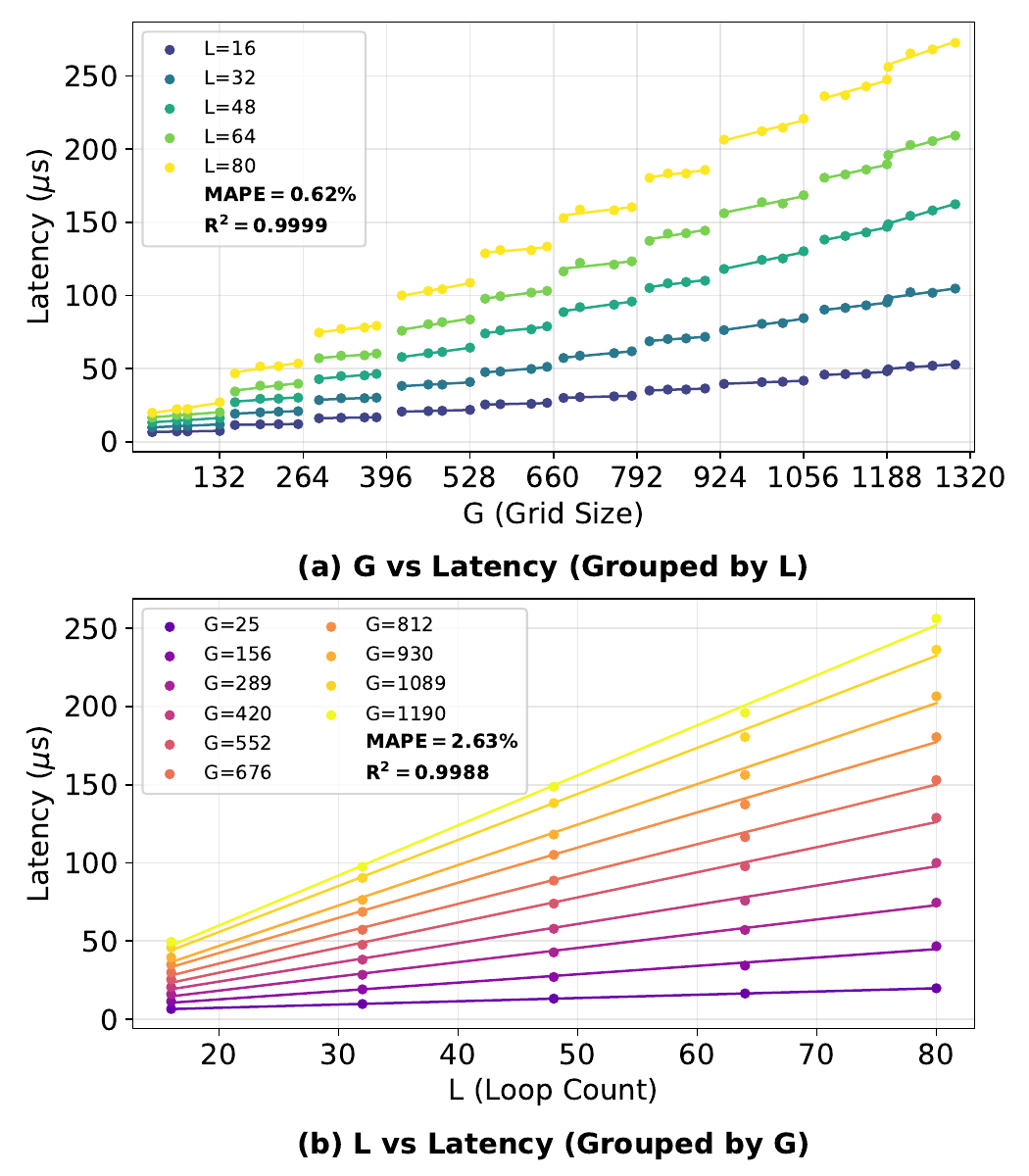}
    \caption{Profiling results on an H100 (132 SMs) for a GEMM kernel with $K=1024 \text{--} 5120$ and tile configuration $(T_M, T_N, T_K) = (128, 64, 64)$. 
    (a) Latency exhibits strong linear scaling with grid size $G$ ($R^2 \approx 0.9999$).
    (b) Latency scales linearly with loop iterations $L$ ($R^2 \approx 0.9988$).}
    \label{fig:gl_linearity}
\end{figure}

\subsection{Piecewise Bilinear Execution Behavior}
\label{obs:linearity}

To bridge the gap between irregular wave dynamics and analytical tractability, we analyze the latency of the GEMM kernel under controlled variables. Despite the microarchitectural variance identified in the previous subsection, we observe a robust structural invariant: the performance profile follows a \textit{piecewise bilinear} behavior.

For executions that yield the same wave count $w$ 
(i.e., $w = \lceil G / N_{SM} \rceil$), the execution duration exhibits dual linearity under a fixed tile configuration, as illustrated in Figure~\ref{fig:gl_linearity}. Specifically, Figure~\ref{fig:gl_linearity}(a) shows that with a fixed loop count ($L$), latency increases linearly with the grid size ($G$). This strict linearity ($R^2 \approx 0.9999$) indicates that execution duration grows proportionally with the aggregate number of thread blocks. Complementarily, Figure~\ref{fig:gl_linearity}(b) reveals that when $G$ is held constant, latency scales linearly with $L$ ($R^2 \approx 0.9988$).

Crucially, the regression lines in Figure~\ref{fig:gl_linearity}(b) are \textit{not parallel}; rather, their slopes increase proportionally with the magnitude of $G$.
This ``fanning out'' effect reveals a fundamental multiplicative interaction between the grid and loop dimensions.
Physically, this interaction is intrinsic to the kernel's workload structure: the aggregate workload volume is proportional to the product of the number of thread blocks and their loop iterations ($GL$).
Therefore, the total execution time is dominated by this aggregate term—whether manifesting as total FLOPs for compute-bound kernels or total memory traffic for memory-bound operations—while secondary overheads manifest as independent marginal costs.

\vspace{0.5em}
\noindent\textit{Consequently, we formalize the kernel execution behavior using a bilinear model:
\begin{equation}
T(G, L) \approx \alpha G L + \beta G + \gamma L + \delta
\label{eq:bilinear}
\end{equation}
Here, $T(G, L)$ denotes the execution latency, and the coefficients $\alpha$, $\beta$, $\gamma$, and $\delta$ are learned from profiling data. The dominant term $\alpha G L$ represents the core execution time driven by the aggregate workload volume, while $\beta G$ and $\gamma L$ model the marginal overheads associated with block scheduling and per-iteration latency. By incorporating the $GL$ interaction, this formulation naturally captures the slope divergence observed in profiling and provides a physically grounded basis for kernel performance prediction.}

\section{Methodology}
To navigate the prohibitively large configuration space of kernel auto-tuning, our methodology is built upon a key observation from Section~\ref{obs_motivation}: 
\textit{GPU kernel latency exhibits a wave-conditioned piecewise bilinear structure,} indicating that the latency surface is not an unstructured black box, but a structured space segmented by wave dynamics. Consequently, exhaustive profiling is unnecessary. Instead, the latency landscape can be effectively approximated by sparsely sampling anchor points within each wave segment and fitting localized linear models.

\begin{figure}[t]
    \centering
    \includegraphics[width=\linewidth]{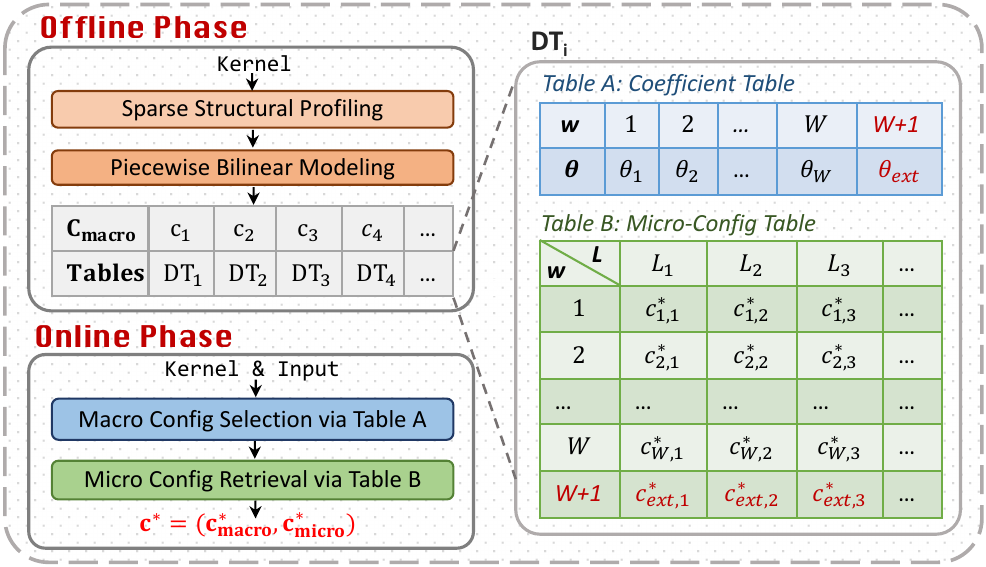}
    \caption{
    Overview of the \textsf{WaveTune} framework.
    For each macro config $c_i$, a corresponding dual-table $\mathrm{DT}(c_i)$ is constructed, consisting of a coefficient table indexed by wave count $w$ with parameters $\boldsymbol{\theta}= \langle \alpha, \beta, \gamma, \delta \rangle$, and a micro-config table indexed by $(w, L)$ storing optimal micro configs $\mathbf{c}^*_{w,L}$. 
    At runtime, Stage I selects $c^*_{\text{macro}}$ via model-based kernel latency prediction, and Stage II retrieves $\mathbf{c}^*_{\text{micro}}$ from the micro-config table.
    }
    \label{fig:WaveTune_overview}
\end{figure}

As illustrated in Figure~\ref{fig:WaveTune_overview}, the overall workflow of \textsf{WaveTune} follows a systematic offline-to-online framework.
First, to address kernel input diversity and configuration complexity, we project complex logical inputs onto a unified 2D physical coordinate system and decouple the configuration space into \textit{macro} and \textit{micro} parameters (Section~\ref{sec:formulation}). 
Next, we perform wave-aware sparse structural profiling to reduce offline data collection overhead (Section~\ref{sec:sparse_structural_profiling}). 
Using collected samples, we construct a dual-table offline artifact: a structural coefficient table parameterizing wave-aware piecewise bilinear models with built-in extrapolation for out-of-range waves, and an anchor-based micro-configuration table for local execution tuning (Section~\ref{sec:wave_modeling} and~\ref{sec:extrapolation_robustness}). 
Finally, at runtime, we apply a two-stage selection process that analytically identifies the optimal macro-config via kernel latency prediction and retrieves micro-config via proximal anchors, enabling deterministic, microsecond-level tuning without expensive hardware evaluations (Section~\ref{sec:two_stage_selection}).

\subsection{Problem Formulation and Parameter Decoupling}
\label{sec:formulation}

We formulate the kernel auto-tuning problem as finding the optimal configuration $\mathbf{c}^*$ that minimizes the execution latency $T$ for a given kernel $\mathcal{K}$ and input shape $\mathbf{x}$:
\begin{equation}
    \mathbf{c}^*(\mathbf{x}) = \operatorname*{argmin}_{\mathbf{c} \in \mathcal{C}} \, T(\mathcal{K}, \mathbf{x}, \mathbf{c})
\end{equation}
Directly solving this optimization is challenging due to the combinatorial explosion of both the logical input space $\mathcal{X}$ and the configuration space $\mathcal{C}$. 
For example, in \texttt{DeepGEMM}, even under commonly used input shapes, the joint space of input dimensions and configuration choices can reach a scale on the order of $10^{15}$ possible combinations. 
To make this tractable, we employ two complementary abstraction strategies.
% We formulate the kernel auto-tuning problem as finding the optimal config $\mathbf{c}^*$ that minimizes the latency $T(\cdot)$ for a given kernel $\mathcal{K}$ and input shape $\mathbf{x}$:
% \begin{equation}
%     \mathbf{c}^*(\mathbf{x}) = \operatorname*{argmin}_{\mathbf{c} \in \mathcal{C}} T(\mathcal{K}, \mathbf{x}, \mathbf{c})
% \end{equation}
% Directly solving this optimization is challenging due to the high dimensionality of both the logical input space $\mathcal{X}$ and the configuration space $\mathcal{C}$. To make this tractable, we employ two complementary abstraction strategies.

\noindent\textbf{Input Mapping: From Logical Inputs to Physical Coordinates.}
Directly modeling kernel latency over original input dimensions is challenging, as these dimensions are kernel-specific and do not directly correspond to hardware execution behavior.
We project the various and variable logical input $\mathbf{x}$ onto a unified 2D physical coordinate system: \textit{Grid Size ($G$) and Loop Count ($L$)}. 
Physically, $G$ represents the spatial partitioning of the workload, determining the total number of logical tasks and thereby governing the wave dynamics. 
In contrast, $L$ captures the temporal workload per task, defined by the iteration count along the reduction axis.
As detailed in Table~\ref{tab:kernel_mapping}, this mapping $\Phi: \mathcal{X} \to (G, L)$ transforms various kernel-specific 
input shapes (e.g., MoE token counts, Attention sequence lengths) into a hardware-oriented canonical representation, reducing the complex logic space to purely physical dimensions.

\noindent\textbf{Configuration Decomposition: Macro and Micro.} 
Simultaneously, we decompose the vast configuration space $\mathcal{C}$ into two functional subspaces: $\mathcal{C} = \mathcal{C}_{\text{macro}} \times \mathcal{C}_{\text{micro}}$. A specific config is therefore expressed as
$\mathbf{c} = (\mathbf{c}_{\text{macro}}, \mathbf{c}_{\text{micro}})$. 
Macro-configs ($\mathbf{c}_{\text{macro}}$) define the \textit{workload granularity} through geometric partitioning, and are intrinsic to the workload geometry, typically following a common partitioning logic across implementations of the same kernel (e.g., tiling M/N/K for GEMM or tiling Q/KV for Attention). This directly determines the physical coordinates $G$ and $L$, thereby governing the global wave dynamics and the dominant resource occupancy characteristics on the GPU.
In contrast, micro-configs ($\mathbf{c}_{\text{micro}}$) optimize the \textit{execution efficiency} of each partition and are often implementation- and hardware-specific. This category encompasses parameters for maximizing instruction-level and thread-level parallelism, such as tuning software pipeline depth for latency hiding, adjusting warp allocation, and shaping thread block rasterization patterns to enhance L2 cache locality. 
Although these two subspaces are coupled via hardware resource constraints—where large macro-tiles may restrict the feasible region of micro-parameters (e.g., shared memory limits)—this decomposition enables structured reasoning.
Table~\ref{tab:kernel_mapping} summarizes representative parameters in each category.
% where we present a subset of representative micro-configs rather than an exhaustive list.
% Macro-configs are intrinsic to the workload geometry and typically follow a common partitioning logic across implementations of the same kernel (e.g., tiling M/N/K for GEMM or tiling Q/KV for Attention). 
% In contrast, micro-configs are often implementation- and hardware-specific. Thus, the table presents a subset of representative micro-parameters rather than an exhaustive list.

\begin{table*}[t]
\centering
\caption{Unified Mapping of Representative Kernels to Physical Dimensions ($G, L$) and Configuration Categories.}
\vspace{-3mm}
\label{tab:kernel_mapping}
\resizebox{\textwidth}{!}{%
\begin{tabular}{@{}lccccc@{}}
\toprule
\textbf{Kernel Type} & \textbf{Logical Inputs ($\mathcal{X}$)} & \textbf{\begin{tabular}[c]{@{}c@{}}Macro Configs ($\mathbf{c}_{\text{macro}}$) \end{tabular}} & \textbf{\begin{tabular}[c]{@{}c@{}}Micro Configs ($\mathbf{c}_{\text{micro}}$) \end{tabular}} & \textbf{Grid Size ($G$)} & \textbf{Loop Count ($L$)} \\ \midrule
\textbf{Dense GEMM} & $M, N, K$ & $T_M, T_N, T_K$ & $N_{stg}, N_{wrp}, \dots$ & $\lceil \frac{M}{T_M} \rceil \times \lceil \frac{N}{T_N} \rceil$ & $\lceil \frac{K}{T_K} \rceil$ \\ \midrule
\textbf{Grouped GEMM (MoE)} & \begin{tabular}[c]{@{}c@{}}$M_i, N, K, E$\end{tabular} & $T_M, T_N, T_K$ & $N_{stg}, N_{wrp}, \dots$ & $\sum_{i=1}^{E} \left( \lceil \frac{M_i}{T_M} \rceil \times \lceil \frac{N}{T_N} \rceil \right)$ & $\lceil \frac{K}{T_K} \rceil$ \\ \midrule
\textbf{FlashAttention} & \begin{tabular}[c]{@{}c@{}}$N_{h}, S_q, S_{kv}$\end{tabular} & $T_Q, T_{KV}$ & $N_{stg}, N_{wrp}, \dots$ & $N_{h} \times \lceil \frac{S_q}{T_Q} \rceil$ & $\lceil \frac{S_{kv}}{T_{KV}} \rceil$ \\ \bottomrule
\end{tabular}%
}
\vspace{1mm}
\footnotesize{\textbf{Note:} $M_i$: Number of tokens routed to the $i$-th expert; $E$: Number of experts; $N_h$: Number of heads; $S_q, S_{kv}$: Sequence lengths; $N_{stg}$: Pipeline stages; $N_{wrp}$: Number of warps.}
\vspace{-4mm}
\end{table*}

\subsection{Sparse Structural Profiling}
\label{sec:sparse_structural_profiling}

Building on the projection $\Phi:\mathcal{X}\rightarrow(G,L)$ and the decoupled configuration space
$\mathcal{C} = \mathcal{C}_{\text{macro}} \times \mathcal{C}_{\text{micro}}$, we next construct a profiling dataset that is both
% $\mathbf{c}=\{\mathbf{c}_{\text{macro}},\mathbf{c}_{\text{micro}}\}$, 
\textit{structurally representative} and \textit{computationally efficient}.
A dense sweep over all feasible tuples $\langle G,L,\mathbf{c}_{\text{macro}},\mathbf{c}_{\text{micro}} \rangle$ is prohibitive;
instead, we perform sparse sampling on the $\langle G,L \rangle$ plane while preserving distinct wave-count execution regimes.

% Let $N_{SM}$ denote the number of SMs, $W$ the profiled wave depth, and $I$ the number of sub-intervals per wave.
\noindent\textbf{Wave-aware sampling of grid size with loop anchors.}
Let $N_{SM}$ denote the number of SMs, $W$ the maximum profiled wave count, and $I$ the number of sub-intervals per wave.
We partition the grid-size axis into wave-aligned regions:
\begin{equation}
\mathcal{R}_w=[(w-1)N_{SM}+1,\;wN_{SM}],\quad w=1,\dots,W,
\end{equation}
and further split each $\mathcal{R}_w$ into $I$ sub-intervals $\mathcal{R}_{w,i}=[a_{w,i},b_{w,i}]$.
From each $\mathcal{R}_{w,i}$, we select a representative grid size $g_{w,i}$ for profiling, forming the sampling set $\mathcal{G}$:
\begin{equation}
\mathcal{G}=\{g_{w,i}\mid w=1,\dots,W,\;i=1,\dots,I\}.
\end{equation}

The representative grid size $g_{w,i}$ is selected according to the kernel structure.
For GEMM-like kernels (e.g., Dense GEMM and Grouped GEMM), we choose the largest value in $\mathcal{R}_{w,i}$ that can be factored into a valid 2D block grid $\langle m_G, n_G \rangle$. Specifically, we require $g_{w,i}=m_G n_G$ ($m_G\le n_G$) while satisfying the aspect-ratio constraint $\rho(g_{w,i})=n_G/m_G\le\tau$ (a predefined upper bound limiting grid elongation, e.g., $\tau=1.1$). This strictly avoids pathological skinny matrices that may introduce atypical memory behaviors. For attention kernels, the grid dimension must align with the head-parallel launch structure.
Thus, we project the interval to the nearest head-aligned value
$g_{w,i}=\left\lfloor b_{w,i}/N_h \right\rfloor \cdot N_h$ with $g_{w,i}\ge a_{w,i}$,
where $N_h$ denotes the number of attention heads.

To unify the sampling rules across different kernels, we adopt simplified kernel abstractions.
Attention kernels are treated as standard multi-head attention (MHA)~\cite{transformer}, since grouped-query attention (GQA)~\cite{GQA} mainly reduces KV-cache storage without changing the dominant compute and memory-access patterns.
Similarly, grouped GEMM is reduced to the single-group case (i.e., dense GEMM), as grouping alters batching structure but not the fundamental behavior.

We further introduce a set of loop anchors to capture different loop behaviors.
Let $\mathcal{L}$ denote the set of sampled loop counts.
The resulting profiling points are the pairs
$\langle G,L \rangle $ with $G\in\mathcal{G}$ and $L\in\mathcal{L}$.
This design covers a wide range of grid sizes and loop counts
while keeping the profiling budget explicitly controlled by the
parameters $(W, I)$ and the number of loop anchors.

\noindent\textbf{From sampled $\langle G,L \rangle $ pairs to executable workloads.}
We generate concrete kernel workloads from the sampled $\langle G,L \rangle $ pairs.
For GEMM-like kernels, the sampled $G$ corresponds to the total number of thread blocks, which has already been factored into the 2D grid dimensions $\langle m_G, n_G \rangle $ during the sampling phase.
Given a macro config $\mathbf{c}_{\text{macro}}=\langle T_M,T_N,T_K \rangle $, the concrete workload dimensions are directly instantiated as:
\begin{equation}
M=m_G T_M,\quad N=n_G T_N,\quad K=L\,T_K .
\end{equation}

For attention kernels, the grid size $G$ corresponds to the number of thread
blocks distributed across attention heads and query tiles.
Under the head-parallel launch structure, each head processes
$G/N_h$ query tiles.
Given a macro config $\mathbf{c}_{\text{macro}}=\langle T_Q,T_{KV} \rangle $, the workload dimensions become
\begin{equation}
S_q=\frac{G}{N_h}T_Q,\quad
S_{kv}=L\,T_{KV}.
\end{equation}
Despite different kernel semantics, both cases follow the same
$\langle G,L \rangle $ abstraction, enabling a unified profiling framework
across heterogeneous kernels.

\noindent\textbf{Micro-config evaluation under the fixed structure.}
For each structural anchor $\langle G,L,\mathbf{c}_{\text{macro}} \rangle$, we evaluate all feasible
$\mathbf{c}_{\text{micro}}\in\mathcal{C}_{\text{micro}}$ using PyTorch Profiler \cite{torch_profiler} as the unified data collection tool.
For each candidate config, we run 3 warm-up iterations followed by 5 measured iterations.

\noindent\textbf{Profiling complexity and output.}
The total number of profiling samples $N_{\text{profile}}$ is
\begin{equation}
N_{\text{profile}}
=
|\mathcal{G}|
\cdot
|\mathcal{L}|
\cdot
|\mathcal{C}_{\text{macro}}|
\cdot
|\mathcal{C}_{\text{micro}}|.
\end{equation}
The resulting dataset records
$\langle G,L,w,\mathbf{c}_{\text{macro}},\mathbf{c}_{\text{micro}},\bar{T} \rangle $,
where $\bar{T}$ denotes the average measured kernel latency.

\subsection{Wave-aware Piecewise Bilinear Modeling}
\label{sec:wave_modeling}

After sparse structural profiling, each sample is represented in the physical space by $\langle G,L \rangle $, together with its measured latency under a feasible macro--micro config pair.
A key observation in Section~\ref{obs_motivation} is that latency is not globally smooth along the $G$-axis: when $G$ crosses wave boundaries, the effective occupancy regime changes, leading to clear phase shifts in the latency surface.
% To preserve this pattern, we adopt a structure-based modeling strategy using $\langle \mathbf{c}_{\text{macro}}, w \rangle $-conditioned buckets, rather than fitting a single global regressor.

To preserve this pattern, we partition the profiling data by macro config $\mathbf{c}_{\text{macro}}$ and wave count $w$, yielding disjoint buckets $\langle \mathbf{c}_{\text{macro}}, w \rangle $, each corresponding to a relatively stable execution regime.
This partitioning reduces cross-regime interference and simplifies local modeling.
Within each $\langle \mathbf{c}_{\text{macro}}, w \rangle$ bucket, we further organize data by loop count $L$.
For each fixed $L$, the bucket contains $I$ sampled grid sizes (one from each sub-interval), each associated with potentially different optimal micro configurations.
Instead of selecting per-point optimal micro configs, we choose a shared $\mathbf{c}_{\text{micro}}^{*}(L)$ that minimizes the average latency across all sampled grid sizes in this $\langle \mathbf{c}_{\text{macro}}, w, L \rangle $ group, and consistently use its measured latency values.
Crucially, these selected $\mathbf{c}_{\text{micro}}^{*}(L)$ are persistently cached as sparse anchor points in a lookup table to serve the runtime micro-retrieval phase (Section~\ref{sec:two_stage_selection}).
This design enforces coherent micro behavior across $G$, allowing the bilinear model to capture pure structural trends over $\langle G,L \rangle $ while avoiding overfitting to pointwise micro-level variations under sparse sampling.

For each $\langle \mathbf{c}_{\text{macro}}, w \rangle $ bucket, we fit a compact bilinear latency model:
\begin{equation}
\hat{T}(G,L \mid \boldsymbol{\theta}_{\mathbf{c}_{\text{macro}}, w})
= \alpha\,GL + \beta\,G + \gamma\,L + \delta,
\label{eq:bilinear_bucket}
\end{equation}
where $\boldsymbol{\theta}_{\mathbf{c}_{\text{macro}}, w} = \langle \alpha, \beta, \gamma, \delta \rangle $ denotes the parameter tuple strictly tied to the given macro-config and wave count.
This formulation provides a highly effective physical abstraction: the cross term $GL$ captures the spatial-temporal coupling between the workload scale and loop iterations, while the linear terms $G$ and $L$ model the fundamental wave scaling and loop execution baselines.
Compared to higher-capacity data-driven regressors (e.g., decision trees and gradient-boosted trees such as XGBoost), this model remains mathematically lightweight, incurs negligible inference overhead, and is highly stable under sparse profiling samples, making it particularly suitable for online kernel selection.

Ultimately, the output of this modeling stage is two lightweight, decoupled lookup tables:
(1) \textbf{Coefficient Table}: Indexed by $\langle \mathbf{c}_{\text{macro}}, w \rangle $, this table stores the bilinear parameter tuples $\boldsymbol{\theta}_{\mathbf{c}_{\text{macro}}, w}$ used for structural latency prediction.
(2) \textbf{Anchor-based Micro-config Table}: Indexed by $\langle \mathbf{c}_{\text{macro}}, w, L \rangle $, this table caches the shared optimal micro-configs $\mathbf{c}_{\text{micro}}^{*}(L)$ identified during the profiling phase.
As shown in Figure \ref{fig:WaveTune_overview}, these tables form the backbone for runtime kernel configuration.
% These two artifacts form the backbone of our runtime execution.
% Rather than relying on complex online computations, the subsequent two-stage configuration selection (Section~\ref{sec:two_stage_selection}) queries this dual-table structure.
% This table-driven design enables deterministic, microsecond-level decision making, translating offline profiling results into low-latency online inference.

\subsection{Two-Stage Runtime Kernel Configuration}
\label{sec:two_stage_selection}

With these two decoupled lookup tables in place, we perform runtime kernel configuration in a two-stage process that directly mirrors the macro/micro decomposition introduced in Section~\ref{sec:formulation}.
The core design principle is to separate \textit{global structural optimization} (macro) from \textit{local execution tuning} (micro), thereby avoiding a prohibitively expensive combinatorial search over the vast space $\mathcal{C}$.

\noindent\textbf{Stage I: Macro selection via table-driven latency prediction.}
Given a runtime input $\mathbf{x}$, we evaluate kernel latency for each candidate macro-config $\mathbf{c}_{\text{macro}} \in \mathcal{C}_{\text{macro}}$.
First, we map the input to its physical coordinates $\langle G, L \rangle$ under each candidate macro-config $\mathbf{c}_{\text{macro}}$, and determine the resulting wave count $w$ induced by the $\mathbf{c}_{\text{macro}}$.
We then retrieve the corresponding parameter tuple $\boldsymbol{\theta}_{\mathbf{c}_{\text{macro}}, w}$ from the table and select the macro-config that minimizes the predicted latency:
\begin{equation}
\mathbf{c}_{\text{macro}}^{*}
=
\arg\min_{\mathclap{\mathbf{c}_{\text{macro}} \in \mathcal{C}_{\text{macro}}}} \,
\hat{T}(G, L \mid \boldsymbol{\theta}_{\mathbf{c}_{\text{macro}}, w}).
\end{equation}
This stage rapidly identifies the optimal workload structure by evaluating the lightweight analytical model, effectively capturing the dominant wave dynamics.

\noindent\textbf{Stage II: Micro retrieval via proximal loop anchors.}
Once the optimal structure $\mathbf{c}_{\text{macro}}^{*}$ is fixed, its corresponding physical dimensions $G^*$ and $L^*$, as well as its wave count $w^*$, are firmly established. We then determine the optimal execution parameters $\mathbf{c}_{\text{micro}}^{*} \in \mathcal{C}_{\text{micro}}$ through a fast lookup to form the final config $\mathbf{c}^* = \langle \mathbf{c}_{\text{macro}}^*, \mathbf{c}_{\text{micro}}^* \rangle $.
As established during the profiling phase, the selected bucket maintains a lookup table of the shared optimal micro-configs $\mathbf{c}_{\text{micro}}^{*}(L)$, pre-computed at sparse loop anchors $L_i$.
For the runtime loop count $L^*$, we perform a \textit{nearest-neighbor retrieval} on the stored anchor set $\mathcal{L}_{\mathbf{c}_{\text{macro}}^{*},w^*}$:
\begin{equation}
\tilde{L}
=
\arg\min_{\mathclap{L_i\in \mathcal{L}_{\mathbf{c}_{\text{macro}}^{*},w^*}}} \,
|L^* - L|.
\end{equation}
The system then directly assigns the cached optimal micro config associated with this proximal anchor, yielding $\mathbf{c}_{\text{micro}}^{*}$ from the anchor indexed by $\tilde{L}$.
This \textit{proximal retrieval} strategy exploits the locality of optimization---where optimal pipeline depths and warp allocations tend to be stable across neighboring loop counts---providing a robust mechanism to handle unseen loop values at negligible overhead.

\noindent\textbf{Why two-stage works.}
This hierarchical approach reduces the runtime parameter configuration complexity from multiplicative ($O(|\mathcal{C}_{\text{macro}}| \cdot |\mathcal{C}_{\text{micro}}|)$) to additive ($O(|\mathcal{C}_{\text{macro}}| + \log |\mathcal{L}|)$). Physically, it aligns with the hardware hierarchy: macro parameters primarily determine workload partitioning and wave structure, while micro parameters mainly refine intra-tile execution efficiency.
By combining analytical latency prediction using table-stored coefficients for macro-level selection with proximal anchor retrieval for micro-level tuning, our approach balances modeling fidelity with low runtime overhead.

\subsection{Extrapolation for Robustness}
\label{sec:extrapolation_robustness}

Although sparse structural profiling covers the dominant operating regions, runtime workloads may still fall outside the profiled wave range.
To handle this case, we augment each macro-config with an \textit{extrapolation model}.
Specifically, besides the per-wave coefficient tuples, we merge samples from the last $p$ (e.g., $p=10$) observed waves and extract an additional set of extrapolation coefficients $\boldsymbol{\theta}^{\text{ext}}_{\mathbf{c}_{\text{macro}}}$ in the same $\langle G,L\rangle$ space, along with a corresponding anchor-based micro-config table constructed from these waves. 
This design leverages the \textit{asymptotic linearity} identified in Section~\ref{obs:wave_dynamics}, where latency was observed to scale linearly with $G$ in the saturation regime.
This provides a stable continuation of the latency trend when the runtime wave index exceeds the profiled horizon.

At inference time, the predictor follows a simple rule: it uses the wave-local coefficient tuple when the queried wave $w$ is inside the profiled range, and switches to the extrapolation coefficients otherwise.
Formally:
\begin{equation}
\hat{T}_{\mathbf{c}_{\text{macro}}}(G,L,w)=
\begin{cases}
\hat{T}(G,L \mid \boldsymbol{\theta}_{\mathbf{c}_{\text{macro}}, w}), & w \le W,\\[1ex]
\hat{T}(G,L \mid \boldsymbol{\theta}^{\text{ext}}_{\mathbf{c}_{\text{macro}}}), & w > W.
\end{cases}
\end{equation}
Correspondingly, the micro-config is retrieved from the micro-config table associated with the selected regime (profiled or extrapolated), preserving the same low-overhead proximal retrieval path detailed in Section~\ref{sec:two_stage_selection}.

Our robustness strategy is therefore twofold:
(1) \textit{structural robustness}, achieved by wave-indexed analytical modeling and asymptotic extrapolation to handle unseen wave regimes; and
(2) \textit{local robustness}, achieved by anchor-based micro retrieval, making the system resilient to moderate loop-depth shifts and profiling sparsity.
Together, these dual mechanisms guarantee prediction continuity and configuration stability without introducing expensive online search.

\section{Evaluation}
\label{evaluation}

\subsection{Experiment Setup}
\label{sec:details}

\noindent\textbf{Kernel Benchmarks.}
We evaluate three representative GPU kernels critical to LLM inference workloads:
(1) \textit{Dense GEMM}: BF16 kernels from \texttt{DeepGEMM}~\cite{deepgemm}, featuring two variants tailored for NVIDIA Hopper and Blackwell architectures.
(2) \textit{Grouped GEMM}: The Triton-based Mixture-of-Experts (MoE) backend from SGLang~\cite{sglang_git}.
(3) \textit{FlashAttention}: The Triton attention backend from vLLM~\cite{vllm_git,triton_attn}, serving as the default implementation for AMD GPUs. These kernels are selected because they dominate the computational cost of LLM inference, accounting for the majority (often over 80\%) of end-to-end execution time~\cite{synperf,vllm_git,sglang_git}.
In contrast to lightweight operators (e.g., RMSNorm or element-wise activations) that have trivial configuration spaces and negligible tuning gains, GEMM and Attention expose massive configuration spaces spanning both macro- and micro-parameters. With the number of valid combinations typically ranging from 200 to over 3,000 per kernel, this combinatorial complexity makes these kernels the primary performance bottlenecks and the most challenging targets for auto-tuning.

\noindent\textbf{Hardware Platforms.}
We evaluate on five modern data-center GPUs from both NVIDIA and AMD: A100, H20, B200, MI308X, and MI355X. This setup enables intra- and cross-vendor comparisons, assessing the robustness of runtime kernel parameter configuration across architectures.

\noindent\textbf{Software Environment.}
NVIDIA experiments utilize CUDA~12.8 on A100 and H20, and CUDA~13.0 on B200, while AMD experiments employ ROCm~6.4.3 on MI308X and ROCm~7.1.1 on MI355X.  
A100, H20, and B200 use PyTorch~2.9.1 with Triton~3.5.1; MI308X runs PyTorch~2.8.0 with Triton~3.4.0; and MI355X uses PyTorch~2.9.0 with Triton~3.4.0.

\noindent\textbf{Baselines.}
We compare \textsf{WaveTune} against four representative runtime configuration baselines.
(1) \textit{Default Heuristics}: The static, human-designed configuration rules hardcoded in the original kernel implementations or framework source code~\cite{deepgemm,sglang_git,vllm_git}.
(2) \textit{Learned Cost Model (GBDT-based)}: A gradient-boosted decision tree model (XGBoost \cite{xgboost}) trained as a surrogate cost model to predict kernel latency, representing a widely used paradigm for accelerating auto-tuning~\cite{autotvm,ansor}.
(3) \textit{Learned Dispatch Policy (Tree-based)}: A lightweight regression tree (Decision Tree) trained to directly map input shapes to configs, approximating data-driven dispatch logic commonly used in practice~\cite{triton_attn}.
(4) \textit{Brute-Force Oracle}: An exhaustive offline auto-tuning search that evaluates the entire configuration space for each input shape, serving as an empirical oracle for the best achievable kernel performance.
To ensure a fair comparison, all learning-based baselines are trained on the same dataset used for training our model.

\noindent\textbf{Dataset Construction.}
We construct comprehensive training and testing datasets to systematically evaluate both modeling accuracy and generalization across diverse workloads. All kernel execution measurements are collected using the PyTorch Profiler \cite{torch_profiler}.

\noindent\textit{Training Dataset.}
For Dense GEMM and Grouped GEMM, we configure $W=40$, $I=4$, and the aspect-ratio upper bound $\tau=1.1$, generating 160 grid points for each of five $K$-dimension anchors spanning $[1024, 5120]$.
For FlashAttention, we adapt the evaluated wave range $W \in [20, 35]$ according to the query head count $qhead \in [16, 128]$, where larger head counts correspond to larger wave ranges. With $I=3$ sub-intervals per wave, this yields 60--105 grid points per KV-length anchor, where KV sequence lengths are scaled by multipliers in the range $[1.0, 3.0]$ to induce varying intra-block loop counts $L$.
Profiling is performed independently on each hardware platform, where we collect approximately 74K configurations for Dense GEMM, 73K for Grouped GEMM, and 6.6K for FlashAttention. The data collection process takes approximately 0.5 to 2 GPU hours per $\langle$hardware, kernel$\rangle$ pair, depending on the workload scale and configuration space size.

\noindent\textit{Testing Dataset.}
To evaluate generalizability, our testing datasets cover extensive input regimes distinct from the training dataset.

\begin{itemize}[leftmargin=*,nosep]
    \item \textbf{Dense GEMM:} 26K samples. $M \in [64, 16384]$, $N \in [384, 8192]$, $K \in [2048, 8192]$.

    \item \textbf{FlashAttention:} 2.8K samples. $seqlen \in [64, 16384]$, $qhead \in [16, 128]$, $kvhead \in [2, 16]$, with a fixed head dimension of 128.

    \item \textbf{Grouped GEMM:} 7.1K samples. $seqlen \in [64, 16384]$, $E \in [8, 32]$, $topk \in [6, 10]$, $H \in [1280, 7168]$, and $N \in [768, 2560]$.
\end{itemize}
To obtain the brute-force oracle, we perform exhaustive auto-tuning over the testing dataset for each $\langle$hardware, kernel$\rangle$ pair. This process is prohibitively expensive and impractical for online deployment, requiring approximately 10 to 600 GPU hours depending on workload scale, kernel complexity, and hardware characteristics.

\noindent\textbf{Model Training.}
We configure both the learning-based baselines and our proposed analytical model using the collected profiling datasets.
The Decision Tree is trained to directly predict the optimal config index from workload features. We use a maximum depth of 15, minimum samples per split of 30, and minimum samples per leaf of 15.
XGBoost is trained as a regression-based cost model to predict kernel latency given both workload and configuration features, using 600 estimators with maximum depth 10, learning rate 0.05, and subsample ratio 0.8; at inference, it evaluates all candidate configs and selects the one with minimum predicted latency. 
\textsf{WaveTune} fits a wave-conditioned analytical latency model using least-squares regression. Separate models are trained per macro-config and wave interval, and inference is performed via constant-time coefficient lookup. All models are trained independently per device to capture hardware-specific performance characteristics. Averaged over each $\langle$hardware, kernel$\rangle$ pair, \textsf{WaveTune} requires only 0.09~MB, compared to 15.25~MB for the Decision Tree and 38.78~MB for XGBoost, indicating its efficiency for deployment.

\subsection{Kernel-Level Performance}
\label{sec:kernel_speedup}

\begin{figure*}[t]
    \centering
    \includegraphics[width=\linewidth]{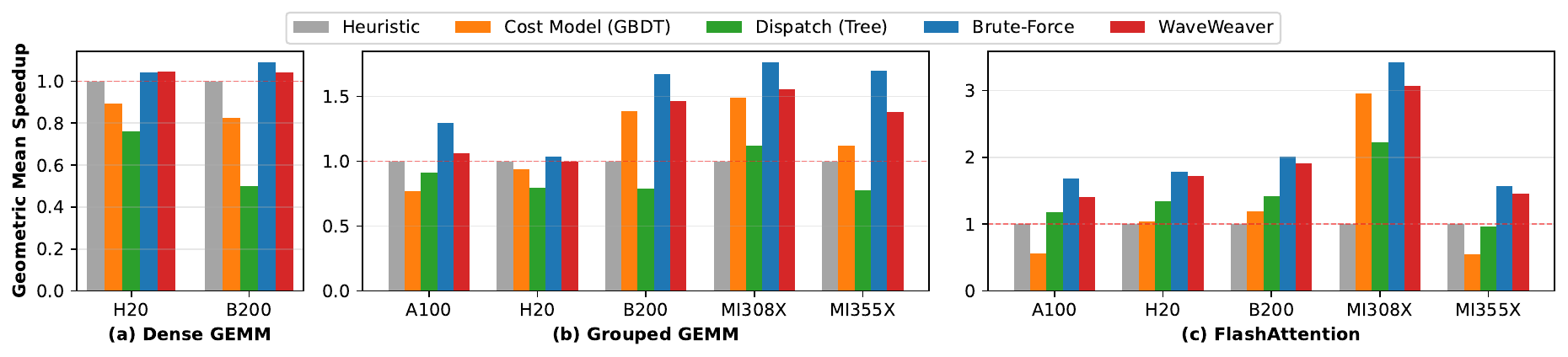}
    \vspace{-8mm}
    \caption{
        Kernel-level geometric mean speedup over the default heuristic across five GPU architectures. \textsf{WaveTune} consistently captures near-optimal performance while avoiding the performance degradations observed in learning-based baselines.
    }
    \label{fig:kernel_performance}
    \vspace{-1mm}
\end{figure*}

\begin{figure*}[t]
    \centering
    \includegraphics[width=\linewidth]{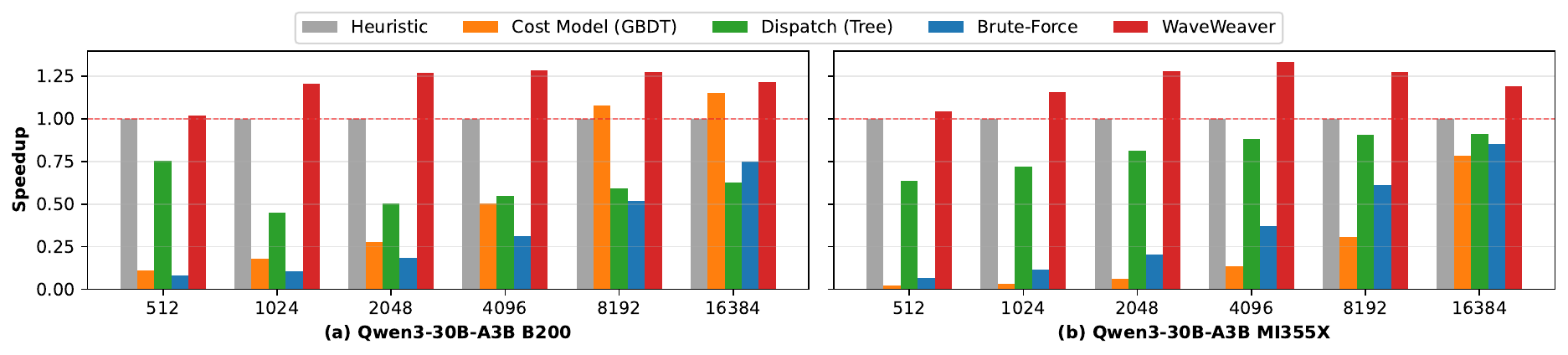}
    \vspace{-8mm}
    \caption{
        End-to-end TTFT speedup over the default heuristic during the prefill phase across varying input sequence lengths. Batch size is fixed at 4.
    }
    \label{fig:e2e_speedup}
    \vspace{-1mm}
\end{figure*}

Figure~\ref{fig:kernel_performance} shows the geometric mean speedup of \textsf{WaveTune} and baselines, normalized to the default heuristic, across three representative kernels and five GPU architectures. Overall, \textsf{WaveTune} consistently delivers the strongest performance among all methods, achieving \textbf{1.83$\times$} speedup on FlashAttention, \textbf{1.27$\times$} on Grouped GEMM and \textbf{1.04$\times$} on Dense GEMM in terms of geometric mean across all hardware platforms, while remaining close to the Oracle upper bounds (2.00$\times$, 1.46$\times$, and 1.07$\times$, respectively).

Learning-based baselines exhibit pronounced cross-kernel instability. The Cost Model shows inconsistent behavior: 1.02$\times$ on FlashAttention, 1.11$\times$ on Grouped GEMM, but dropping to 0.86$\times$ on Dense GEMM. Dispatch degrades even more dramatically, falling from 1.37$\times$ on FlashAttention to 0.87$\times$ on Grouped GEMM and further to 0.62$\times$ on Dense GEMM. In contrast, \textsf{WaveTune} maintains consistent performance across all kernels without any regression, even in optimization-constrained regimes (e.g., Dense GEMM with only 1.07$\times$ Oracle headroom).
% This contrast is most pronounced on the heavily optimized Dense GEMM kernel, where the optimization margin is limited (Oracle: 1.07$\times$). In this constrained regime, both learning-based baselines tend to overfit and incur significant regressions, while \textsf{WaveTune} remains stable with a 1.04$\times$ speedup, demonstrating that wave-based modeling more reliably captures underlying hardware behavior when optimization headroom is limited.

On Grouped GEMM, \textsf{WaveTune} achieves a 1.27$\times$ speedup, with a larger gap to the Oracle (1.46$\times$) than in other kernels. This gap arises from the inherent uncertainty in MoE token routing: tokens are dynamically assigned to experts at runtime, making per-expert token counts unpredictable. Accurately querying these distributions would require synchronization that breaks CPU–GPU asynchrony and incurs prohibitive overhead. As a result, \textsf{WaveTune} adopts a uniform approximation across experts, trading peak optimality for practical deployability. Despite this approximation, it still outperforms all baselines and maintains stable gains across hardware, demonstrating robustness under real-world serving constraints.

% Overall, these results show that \textsf{WaveTune} not only approaches Oracle-level performance but also maintains strong robustness across diverse kernels and hardware platforms, outperforming both heuristic and learning-based baselines in terms of accuracy and stability.

\subsection{End-to-End Inference Performance}
\label{sec:e2e_speedup}

To evaluate the system-level impact of \textsf{WaveTune}, we integrate it into the Triton-based Grouped GEMM kernel for MoE workloads within the SGLang 0.5.8 inference engine. We do not modify the kernel implementation, but only replace the runtime configuration policy in the SGLang framework.

Figure~\ref{fig:e2e_speedup} reports the TTFT speedup of each method, normalized to the default heuristic, on NVIDIA B200 and AMD MI355X. Experiments are conducted for single-GPU inference using the Qwen3-30B-A3B \cite{qwen3_report} model with a fixed batch size of 4, while varying the input sequence length from 512 to 16,384 tokens to stress the prefill phase. We focus on TTFT as it is dominated by the prefill stage, where long input sequences induce large computational workloads and expose substantial configuration tuning opportunities. In contrast, the decode stage operates on very small token counts per step, offering limited parallelism and leaving little room for configuration optimization; in this regime, simple or fixed heuristics are typically sufficient to achieve near-optimal performance.

Across both GPU architectures, \textsf{WaveTune} consistently delivers significant TTFT reductions, achieving up to \textbf{1.33$\times$} speedup on MI355X and a peak \textbf{1.28$\times$} speedup on B200.
These end-to-end results highlight a fundamental trade-off between configuration overhead and execution quality. Exhaustive online search (Brute-Force) incurs prohibitive runtime overhead. Triggering auto-tuning---even when combined with caching of the best configuration—can introduce \textit{second-level} latency due to the need to evaluate a large configuration space. This overhead dominates TTFT at small sequence lengths and makes it fundamentally impractical for latency-sensitive serving. As the sequence length increases, the search cost can be partially amortized over longer execution, but the overall performance remains worse than the default heuristic in end-to-end scenarios.
Similarly, the GBDT-based Cost Model suffers from substantial runtime overhead due to CPU-side inference. This overhead dominates in the low-latency regime, leading to severe slowdowns at small sequence lengths. As workload scales and GPU execution becomes the primary bottleneck, the CPU-side inference cost can be partially masked by asynchronous CPU–GPU execution. As a result, the Cost Model begins to recover on B200 and approaches the performance of \textsf{WaveTune} at 16{,}384 tokens. However, this behavior does not generalize across architectures: on MI355X, even at the largest sequence length, it remains inferior to the default heuristic.
In contrast, the Dispatch baseline introduces relatively low runtime overhead but exhibits consistently suboptimal config choices across different input lengths, resulting in degraded or unstable performance.

\textsf{WaveTune} eliminates this trade-off by replacing expensive runtime search with an efficient model-based prediction mechanism. With lightweight table lookups and bilinear evaluation, it achieves near-zero overhead while retaining near-optimal performance. This enables \textsf{WaveTune} to consistently achieve the lowest TTFT across all sequence lengths, delivering robust gains in real-world LLM serving systems.

\subsection{Runtime Overhead}
\label{sec:overhead}

\begin{figure}[t]
    \centering
    \includegraphics[width=1.01\linewidth]{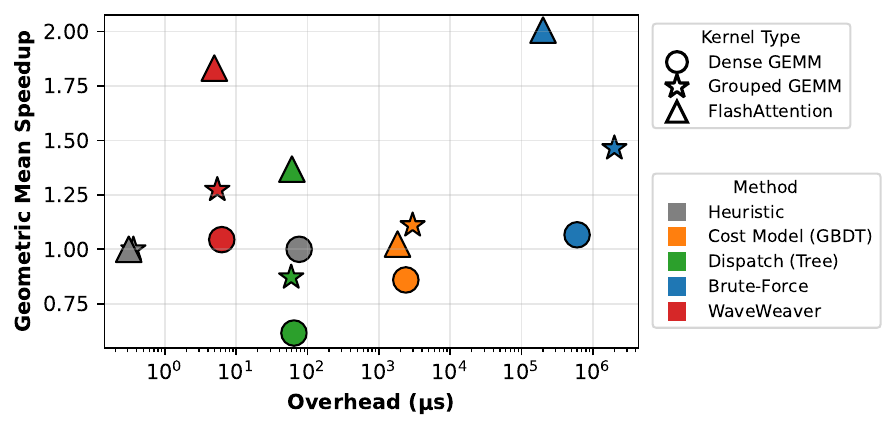}
    \vspace{-7mm}
    \caption{
        Trade-off between runtime decision overhead (log scale) and overall geometric mean speedup. \textsf{WaveTune} lies in the favorable upper-left region.
    }
    \label{fig:overhead_speedup}
    \vspace{-3mm}
\end{figure}

To be viable for production serving, a configuration policy must make deterministic decisions within microseconds to avoid stalling the asynchronous GPU execution pipeline. Figure~\ref{fig:overhead_speedup} quantifies the trade-off between execution speedup and runtime decision overhead, providing a concrete breakdown of the system-level behavior observed in Section~\ref{sec:e2e_speedup}. The GBDT-based Cost Model incurs substantial CPU-side inference overhead due to sequential tree traversal across multiple candidates, resulting in decision latencies ranging from 1,822\,$\mu$s to 2,965\,$\mu$s , which already exceeds the execution time of most kernels. The Dispatch baseline (Decision Tree) reduces this overhead but still introduces around 60\,$\mu$s of branching cost, while failing to consistently select high-quality configs.

In contrast, \textsf{WaveTune} achieves near-oracle execution performance with a tightly bounded runtime overhead of \textbf{5--6\,}$\mu$s across all workloads, enabled by its $O(1)$ dual-table lookup and lightweight bilinear evaluation. For heavily optimized kernels such as \texttt{DeepGEMM}, the default heuristic itself incurs a 76\,$\mu$s overhead due to on-the-fly hardware constraint evaluation, whereas \textsf{WaveTune} reduces this cost to \textbf{6\,}$\mu$s (a \textbf{12$\times$} reduction) while also improving execution latency. Overall, \textsf{WaveTune} effectively eliminates the practical trade-off between configuration quality and decision latency, delivering both low overhead and strong performance consistently.

\subsection{Ablation Study}
\label{sec:ablation}

\begin{figure}[t]
    \centering
    \includegraphics[width=\linewidth]{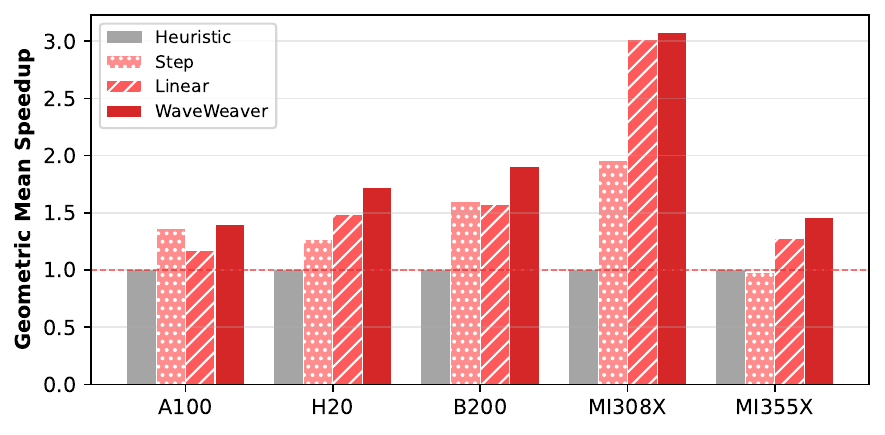}
    \vspace{-7mm}
    \caption{Ablation study on FlashAttention, comparing the full \textsf{WaveTune} model against its Linear and Step variants.}
    \label{fig:ablation_comparison}
    \vspace{-3mm}
\end{figure}

\subsubsection{Effect of Modeling Components.}
To validate the core design of \textsf{WaveTune}'s latency modeling, we conduct an ablation study on the FlashAttention kernel across five GPU architectures, using the same training data and evaluation setup as in the main experiments. We compare the full model against two degraded variants: (i) \textbf{Linear}, which removes wave-aware partitioning and fits a single bilinear model over the entire space, and (ii) \textbf{Step}, which removes intra-wave linear structure and models latency solely as a function of wave count. Figure~\ref{fig:ablation_comparison} reports geometric mean speedups normalized to the default heuristic. 

The full \textsf{WaveTune} model achieves the highest overall speedup of 1.83$\times$, while the Linear and Step variants drop to 1.60$\times$ and 1.39$\times$, respectively. The degradation of the Linear variant indicates that a global bilinear model fails to capture the latency discontinuities at wave boundaries. This limitation is particularly severe in the small-wave regime, where discrete wave effects dominate and the model incurs large errors, leading to poor configuration selection. In contrast, its performance improves at larger wave counts, where the latency surface becomes increasingly smooth and closer to linear, consistent with the asymptotic behavior observed in Section~\ref{obs:wave_dynamics}. As a result, the Linear model exhibits unstable performance across regimes and remains suboptimal overall (e.g., 1.91$\times \rightarrow$ 1.58$\times$ on B200 and 1.72$\times \rightarrow$ 1.48$\times$ on H20). The Step variant performs worse, as it ignores intra-wave scaling and treats all configurations within a wave as equivalent, reducing its ability to differentiate execution efficiency; this even results in a regression on MI355X (0.98$\times$). These results show that neither global continuous modeling nor pure step modeling is sufficient in isolation. Accurate performance modeling requires jointly capturing inter-wave boundaries and intra-wave dynamics.

\subsubsection{Effect of Profiling Range $(W, I)$.}
\begin{figure}[t]
    \centering
    \includegraphics[width=\linewidth]{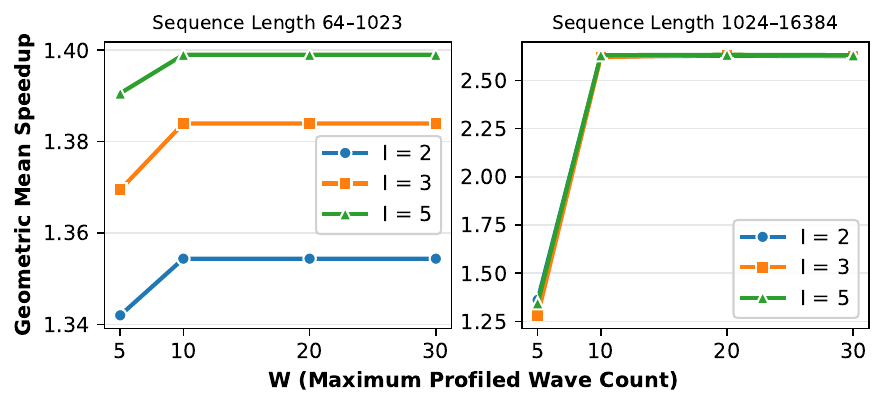}
    \vspace{-7mm}
    \caption{Impact of profiling range $(W, I)$ on performance across different sequence length ranges.}
    \label{fig:data_scaling}
    \vspace{-3mm}
\end{figure}

We further study the impact of the profiling range by varying the maximum wave count $W$ and the number of intervals $I$, as shown in Figure~\ref{fig:data_scaling}. 
The experiments are conducted on NVIDIA B200 with attention head count fixed to 16, while all other settings follow the main evaluation.

For sequence lengths in the range $[64, 1023]$, where the corresponding wave counts $w$ in the test set fall within $[1, 7]$, increasing $I$ consistently improves performance, while increasing $W$ from 5 to 10 also yields gains, but saturates beyond this point. This is because the maximum wave count in the test set is below 10, and thus further expanding the profiling range provides no additional coverage.
In contrast, for sequence lengths in the range $[1024, 16384]$, where $w$ spans a much wider range $[1, 108]$, performance quickly saturates once $W$ reaches 10, and varying $I$ has minimal impact. This is because execution becomes more stable and approaches linear scaling at larger sequence lengths, making small $W$ and coarse $I$ sufficient to capture the dominant performance characteristics.
Importantly, despite the limited profiling range, the model maintains strong performance even when runtime wave counts exceed the profiled horizon, validating the effectiveness of our extrapolation mechanism. In general, these results show that \textsf{WaveTune} achieves high performance with sparse coverage of the $(W, I)$ space.

\section{Related Work}

\noindent\textbf{Auto-Tuning for GPU Kernels.}
Search-based auto-tuning has been widely adopted to optimize GPU kernel performance by exploring large configuration spaces. Early systems such as AutoTVM~\cite{autotvm} and Ansor~\cite{ansor} employ learned cost models (e.g., gradient-boosted trees) to guide search and reduce tuning cost. Subsequent works further improve search efficiency through program sampling and transfer learning~\cite{meta_schedule}. Analytical approaches such as Roller~\cite{roller} attempt to model hardware behavior to prune the search space. However, these methods are primarily designed for offline optimization and still rely on search at deployment time, making them unsuitable for latency-critical online serving scenarios.

\noindent\textbf{Heuristic-based Configuration.}
In production systems, runtime configuration is typically governed by manually designed heuristics embedded in kernel libraries and frameworks~\cite{cublas,cutlass_docs,triton_attn,vllm_git,sglang_git}. These heuristics provide low-latency decisions but are labor-intensive to develop and often fail to generalize across diverse workloads and hardware architectures.

\noindent\textbf{GPU Performance Modeling.}
Prior work on GPU performance modeling includes cycle-accurate simulators~\cite{gpgpusim,accelsim,mgpusim}, analytical models~\cite{AMALI,GCOM,llmcompass}, and data-driven approaches~\cite{habitat,neusight,synperf}. While these methods aim to predict kernel latency or throughput, they typically focus on performance estimation rather than direct configuration. High-fidelity simulators incur prohibitive computational cost, while analytical and learned models often rely on simplified assumptions or introduce non-trivial inference overhead. As a result, these approaches are not well suited for real-time, fine-grained kernel configuration in latency-critical serving scenarios.

% Table~\ref{tab:design_space} summarizes the trade-offs among existing configuration selection strategies. Offline search\cite{hipautotune,benchmarking_autotuning,autotvm} achieves high accuracy but lacks generalization and incurs high runtime cost, while expert heuristics\cite{vllm_git,sglang_git,triton_attn,tritonblas} provide efficient decisions but suffer from poor performance and limited portability. Learning-based cost models\cite{autotvm,ansor,xgboost} partially improve accuracy but introduce non-negligible inference overhead.
% In contrast, \textsf{WaveTune} achieves high accuracy and strong generalization while maintaining constant-time runtime efficiency, effectively unifying these desirable properties within a single framework.

%压缩版本
\section{Conclusion}

We present \textsf{WaveTune}, a runtime kernel configuration framework that resolves the trade-off between execution optimality and decision overhead in dynamic LLM serving, enabling accurate latency prediction with near-zero overhead.
Across three kernels and five GPU architectures, it achieves up to \textbf{1.83$\times$} kernel speedup and \textbf{1.33$\times$} TTFT reduction, while reducing decision overhead by five orders of magnitude over exhaustive search, demonstrating that lightweight wave-aware modeling enables high performance with efficient runtime decision making.

% \section{Conclusion}

% We present \textsf{WaveTune}, a runtime kernel parameter configuration framework that resolves the trade-off between execution optimality and decision overhead in dynamic LLM serving. By leveraging a wave-aware piecewise bilinear model, it enables accurate latency prediction with near-zero overhead. Across three kernels and five GPU architectures, it achieves up to \textbf{1.83$\times$} kernel speedup and \textbf{1.33$\times$} TTFT reduction, while reducing runtime decision overhead by five orders of magnitude compared to exhaustive search, demonstrating that lightweight, wave-aware modeling can simultaneously deliver high performance and low overhead runtime decision making.

% \begin{acks}
% We thank colleagues for helpful discussions. This work used LLMs for text refinement and code generation. We also thank providers of computational resources that supported this work.
% \end{acks}

%%%%%%% -- PAPER CONTENT ENDS -- %%%%%%%%

\bibliographystyle{ACM-Reference-Format}
\bibliography{ref}

\end{document}